\documentclass[fleqn]{eptcs}
\usepackage{amssymb}
\usepackage{amsmath}
\usepackage{graphicx}
\usepackage{stmaryrd}

\newcommand{\vev}[1]{#1}

\newcounter{soscounter}

\newcommand{\pbis}[1]{\mathbin{\preceq_{#1}}}
\newcommand{\bpbis}[1]{\mathbin{\leftrightarrow_{#1}}}

\newcommand{\val}[2]{\langle #1, #2 \rangle}

\newcommand{\boolf}{\setdisp{B}}

\newcommand{\vt}{\mathrm{true}}
\newcommand{\vf}{\mathrm{false}}
\newcommand{\gc}[2]{#1 :\rightarrow\! #2}

\newcommand{\emp}{\st}

\newtheorem{example}{Example}

\newcommand{\Cterms}{\setdisp{T}}

\newcommand{\bbbn}{\mathbb{N}}

\newcommand{\setdisp}[1]{\mathsf{#1}}
\newcommand{\fndisp}[1]{\mathrm{#1}}

\newcommand{\sos}[2]{
    \refstepcounter{soscounter}
    \mbox{\small $\mathbf{\thesoscounter}~$}
    \displaystyle \frac{#1}{#2}
}

\newcommand{\sossym}[2]{
    \refstepcounter{soscounter}
    \mbox{\small $\mathbf{\thesoscounter}~$}
    \refstepcounter{soscounter}
    \mbox{\small $\mathbf{\!(\thesoscounter)}~$}
    \displaystyle \frac{#1}{#2}
}

\newcommand{\pc}{\parallel}

\newcommand{\dl}{0}
\newcommand{\st}{1}

\newcommand{\encap}[2]{\partial_{#1}\left(#2\right)}

\newcommand{\dom}{\fndisp{D}}

\newcommand{\Sdat}{\setdisp{D}}
\newcommand{\Sact}{\setdisp{A}}
\newcommand{\Sexp}{\setdisp{F}}

\newcommand{\Sreq}{\setdisp{R}}
\newcommand{\Svar}{\setdisp{V}}
\newcommand{\Schan}{\setdisp{H}}
\newcommand{\Uact}{\setdisp{U}}
\newcommand{\Cact}{\setdisp{C}}

\newcommand{\tss}[1]{#1\mkern1mu \mathord{\downarrow}}

\newcommand{\tr}[1]{\mathbin{\stackrel{#1}{\mathord{\longrightarrow}}}}

\newcommand{\trs}[1]{\mathbin{\stackrel{{#1}}{\mathord{\longrightarrow}}{\mkern-8mu}^*}}
\newcommand{\ntr}[1]{\mathbin{\stackrel{#1}{\mathord{\mkern4mu \Arrownot \mkern-4mu \longrightarrow}}}}

\title{Communicating Processes with Data for\\Supervisory Coordination}
\author{Jasen Markovski\thanks{Supported by Dutch NWO project: ProThOS, no. 600.065.120.11N124.}
\institute{
Department of Mechanical Engineering,
Eindhoven University of Technology,\\
P.O.~Box~513, 5600 MB Eindhoven, The Netherlands,}
\email{j.markovski@tue.nl}
}
\def\titlerunning{Communicating Processes with Data for Supervisory Coordination}
\def\authorrunning{J. Markovski}

\begin{document}

\maketitle
\begin{abstract}
We employ supervisory controllers to safely coordinate high-level discrete(-event) behavior of distributed components of complex systems. Supervisory controllers observe discrete-event system behavior, make a decision on allowed activities, and communicate the control signals to the involved parties. Models of the supervisory controllers can be automatically synthesized based on formal models of the system components and a formalization of the safe coordination (control) requirements. Based on the obtained models, code generation can be used to implement the supervisory controllers in software, on a PLC, or an embedded (micro)processor.
In this article, we develop a process theory with data that supports a model-based systems engineering framework for supervisory coordination. We employ communication to distinguish between the different flows of information, i.e., observation and supervision, whereas we employ data to specify the coordination requirements more compactly, and to increase the expressivity of the framework. To illustrate the framework, we remodel an industrial case study involving coordination of maintenance procedures of a printing process of a high-tech Oc\'{e} printer.
\end{abstract}

\section{Introduction}

Traditional software development techniques proved insufficiently flexible for development of quality control software, establishing the latter as an important bottleneck in design and production of complex high-tech systems~\cite{controlsoftware}. This gave rise to supervisory control theory of discrete-event systems~\cite{rwsupervisor,Cassandras} that studies automatic synthesis of (discrete-event) models of supervisory control software.

\begin{figure}[!h]
\centering
\begin{tabular}{l@{\hspace{-0.5cm}}ll@{\hspace{-0.5cm}}l}
a) &
\includegraphics[width = 0.65\linewidth]{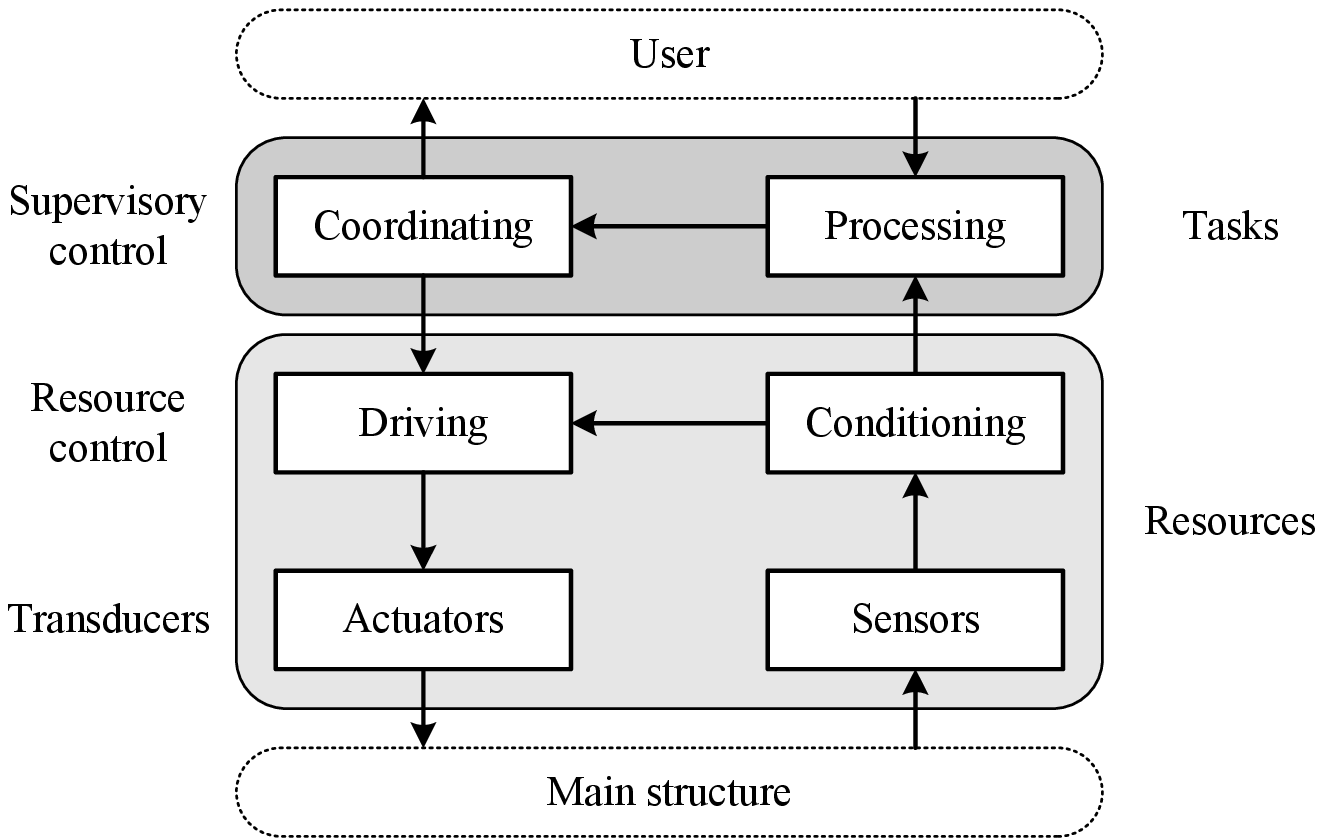} &
b) &
\raisebox{2cm}{
\includegraphics[width = 0.32\linewidth]{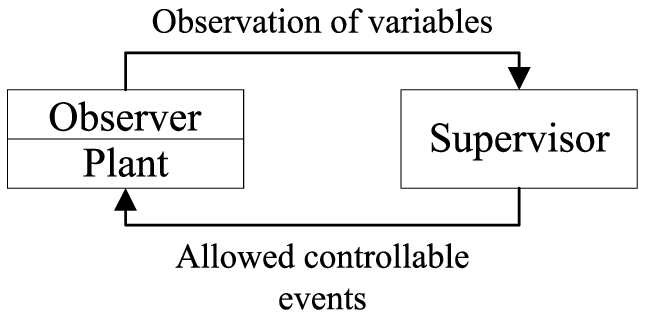}}
\end{tabular}
\caption{a) Supervisory control; b) Supervisory control feedback loop with data-based observations}\label{fig:intro}
\end{figure}

\subsection{Supervisory Control}

Supervisory controllers safely coordinate high-level system behavior, relying on observations made regarding the discrete(-event) system behavior by using sensory information, as depicted in Figure~\ref{fig:intro}a). Based upon the observed signals, the supervisory controllers make a decision on which activities are allowed to be carried out safely, and send back control signals to the hardware actuators. By assuming that the supervisory controller reacts sufficiently fast on machine input, one can model this \emph{supervisory control feedback loop} as a pair of synchronizing processes~\cite{rwsupervisor,Cassandras}. The model of the machine, which is referred to as \emph{plant}, is restricted by the model of the controller, referred to as \emph{supervisor}. The synchronization of the supervisor and the plant, results in the \emph{supervised plant}, which models the supervisory control loop, i.e., it specifies the behavior of the supervised system.

The activities of the machine are modeled as discrete events, whereas the supervisor is a process that synchronizes with the plant, and traditionally, it disables events by not synchronizing with them, whereas it enables events by synchronizing with them~\cite{rwsupervisor,Cassandras}. As a result, the supervisor comprises the complete history of the supervised system, i.e., it enumerates the state space of the supervised system~\cite{Cassandras,acc2011}. The events are split into \emph{controllable events}, which can be disabled by the supervisor in order to prevent potentially dangerous or otherwise undesired behavior, and \emph{uncontrollable events}, which must never be disabled by the supervisor. The former model activities over which control can be exhibited, like interaction with the actuators of the machine, whereas the latter model activities beyond the control of the supervisor, like observation of sensors or user interaction with the environment. Moreover, the supervised plant must also satisfy the \emph{control requirements}, which model the safe or allowed behavior of the machine. In addition, it is typically required that the supervised plant is nonblocking, meaning that it comprises no deadlock and no livelock. To this end, every state is required to be able to reach a so-called \emph{marked} or final state~\cite{rwsupervisor,Cassandras}, which denote states in which the plant is considered to have successfully completed its execution. The conditions that ensure the existence of such a supervisor are referred to as (nonblocking) \emph{controllability} conditions~\cite{rwsupervisor,Cassandras}.

\subsection{Motivation and Contributions}

Our initial motivation for developing a process theory that distinguishes between the different flows of information between the plant and the supervisor is the oversimplification of the modeling of the supervisory control loop in the original proposal of~\cite{rwsupervisor,Cassandras}. This manner of representation of this communication, by means of synchronizing action using automata-style synchronization, still prevails in modern state-of-the-art approaches, like~\cite{heymann-nondeterministic,fabiannondeterministic,overkamp-nondeterministic,kumarnondeterministic}. This is duely noted in~\cite{PACO2011}, where a proposal is given to separate the different flows of information and to give a separate characterization of the process forms of the plant and the supervisor.

The approach investigated in~\cite{PACO2011} relies on propositional signals that stem from the states such that the supervisor has (intrinsic) knowledge regarding the state of the plant. Typically, state- or data-based approaches to supervisory control~\cite{MaWonham,CDC2010,Cassandras} require the use of an \emph{observer}, which represents an addition to the plant as depicted in Figure~\ref{fig:intro}b). The observer derives the state of the plant based on the history of observed events such that it can be directly communicated to the supervisor and employed for supervision. There are a couple of issues in the proposal of~\cite{PACO2011} when attempting to employ the process theory for modeling of supervisory control loops similar to the one depicted in Figure~\ref{fig:intro}b). Namely, the semantics of the propositional signals relies on a predefined nondeterministic valuation effect function that updates the propositional signals based on the label of the taken transition and a set of possible future propositional signals~\cite{PACO2011}. This inevitably leads to unnecessary nondeterministically-chosen deadlock states when the intended propositional signal is not observed, making these deadlock states hard to interpret in a supervisory control setting. In addition, the observation of signals implicitly implies that the supervisor observes the state of the plant, not distinguishing between the plant and the observer. Admittedly, this is a standard practice, especially when modeling complex systems and development of compact and approachable models is of interest. Nonetheless, the underlying process theory should depict these nuances in a subtler manner.

To address the issues outlined above, we propose to replace and extend the propositional signals with variable assignments, which dynamically determine the valuation effect function resolving the first issue. As a welcome side effect, we obtain a compacter set of operational rules than the one presented in~\cite{PACO2011}. Moreover, by not having to implicitly couple the semantics of states with propositional signals, we have the option to model observers either as an intrinsic (integrated) part of the plant or as a separate process. In the setting of this paper, we rely on data-based observations, as depicted in Figure~\ref{fig:intro}b). As discussed, the plant is augmented with an observer process, which may assign auxiliary data variables, based on the history of observed event. These data is required by the supervisor in order to make the correct control decision. We illustrate the situation by an example.

\begin{figure}[!h]
\centering
\includegraphics[width = \linewidth]{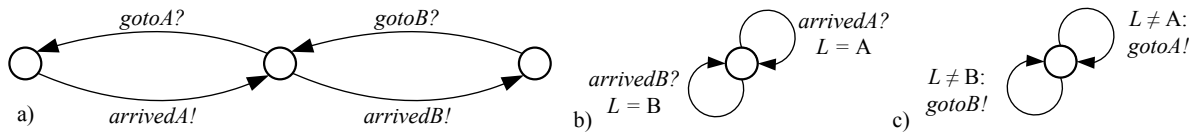}
\caption{a) A plant that models the behavior of an automated guided vehicle; b) An observer that keeps track of the location of the vehicle of a); c) A supervisor that ensures proper coordination of the vehicle of a)}\label{fig:exintro}
\end{figure}

\begin{example}\label{ex:intro}
Let us assume that we have an automated guided vehicle that is capable of traveling to two positions, say A and B. We can issue two commands to the vehicle, namely \emph{gotoA} and \emph{gotoB}, so that the vehicle travels to A and B, respectively. When the vehicle arrives at the corresponding location, it reports back using the sensor signals \emph{arrivedA} and \emph{arrivedB}, respectively. We can model the behavior of this vehicle using the simple transition system, depicted in Figure~\ref{fig:exintro}a). Note that we distinguish between the direction of communicated commands and signals. By \emph{event?} we denote a recipient party of the communication, and by \emph{event!} we denote a sender party. We employ generic communication events, e.g., \emph{event$!_2?_3$} denotes a resulting communication event that occured between two sender and three recipient parties.

Now, suppose that we wish to coordinate the movement of this vehicle, such that if the vehicle is at location $X$, then we do not issue a redundant command that sends the vehicle to the location $X$, for $X \in \{\mathrm{A}, \mathrm{B}\}$. Obviously, by employing only the transition system in Figure~\ref{fig:exintro}a) such coordination is not possible, since the model of the behavior of the vehicle does not comprise information regarding the location of the vehicle. To this end, we need an observer, which updates a variable, say $L$, with respect to the current location of the vehicle, as depicted in Figure~\ref{fig:exintro}b). The observer waits for confirmation signal, sent from the vehicles, that it has reached the corresponding location in order to update its status. By employing this location information, the supervisor can make the correct decision on which command is allowed to be issued.

Now, we can state the coordination or control requirements, based on the data observations as: if $L = \mathrm{X}$, then the event \emph{gotoX} should not be enabled, for $X \in \{\mathrm{A}, \mathrm{B}\}$. By employing these control requirements, we can synthesize the supervisor depicted in Figure~\ref{fig:exintro}c). We note that the supervisor enables the movement commands, relying on guards that observe the shared location variable that is provided by the observer.
\end{example}

To capture the controllability conditions involving the plant, the control requirements, and the supervisor we rely on a behavioral preorder termed \emph{partial bisimulation}. In essence, we employ this preorder to state a relation between the supervised plant and the original plant allowing controllable events to be simulated, while requiring that uncontrollable event are bisimulated. This ensures that the supervisor does not disable uncontrollable events, while preserving the branching structure of the plant. 
Previous proposals like~\cite{heymann-nondeterministic} and~\cite{overkamp-nondeterministic} rely on the process theory CSP~\cite{CSP}, whereas other approaches rely on trace-based notions to capture controllability~\cite{fabiannondeterministic,kumarnondeterministic}. In~\cite{heymann-nondeterministic} the theory is extended with a specialized prioritized communication operator that captures the communication between the plant and the supervisor, later replaced by a refinement relation in failure semantics~\cite{overkamp-nondeterministic}. The control requirements depend on the observed data, and they are given in terms of global invariants that depend on the allowed data assignments, or they specify when a given event is allowed or disallowed, similarly to the informal specification of the control requirements in Example~\ref{ex:intro}.

In the remainder of this paper, we first present the process theory and, thereafter, we discuss its application in a model-based systems engineering framework for supervisory coordination and control. To illustrate the framework, we revisit a case study that deals with coordination of maintenance procedures of a printing process of an Oc\'{e} prototype printer~\cite{wodes2010}. The control problem is to synthesize a supervisory coordinator that ensures that quality of printing is not compromised by timely performing maintenance procedures, while interrupting ongoing print jobs as little as possible. Unlike previous attempts~\cite{PACO2011}, we parameterize the model to handle multiple maintenance procedures concurrently. Due to confidentiality issues, we can only present an (obfuscated) part of the case study.

\section{Communicating Processes with Data}

To model data elements and guards, we extend the process theories $\mathrm{BSP}_{\pc}$ of~\cite{acc2011} and TCP* of~\cite{PACO2011}, thus obtaining communicating processes with data. The result is a process theory encompassing \emph{successful termination} that indicates final or marked states~\cite{rwsupervisor,Cassandras}, which model that the plant can successful terminate its execution; \emph{generic communication action prefixes with data assignments}, which model activities of the plant and enable a dynamic valuation effect function; \emph{guarded commands}, which condition transitions based on data assignments, and enable data observation and support supervision; \emph{sequential composition}, which is an auxiliary operator required for definition of recursive processes; \emph{iteration} to specify recurring behavior; and ACP-style \emph{parallel composition with synchronization}~\cite{rwsupervisor} and \emph{encapsulation}, which model together a flexible coupling in the feedback control loop based on given communication parties. We note that additional process operations can be easily added in the vein of~\cite{pabook,PACO2011}.

We remark that the synthesis tool Supremica~\cite{supremica}, which we employ in the implementation of the model-based system engineering framework, supports the automata-like synchronization of~\cite{rwsupervisor,Cassandras}, which is standardly used in supervisory control theory. Moreover, there exists no distinction between sender and receiver parties in the parallel composition. The automata-like parallel composition synchronizes on all events from all processes that are in the common alphabet, whereas the remainder of the events is interleaved. It is not difficult to show, e.g., in the vein of~\cite{reportPACO}, that our setting subsumes this parallel composition.

\subsection{Syntax}

In principle, we allow data elements to be of any type, given by the set $\Sdat$, even though only finite integer and enumerated types are currently supported by the synthesis tool~\cite{supremica}. By $\Svar$, we denote the set of data variables, and by $\Sexp$, data expressions involving standard arithmetical operations supported by Supremica~\cite{supremica}. The arithmetical operations are evaluated with respect to $e \colon \Sexp \to \Sdat$. The guarded commands are given as Boolean formulas, whereas the atomic propositions are given by the predicates from the set $\{<, \le, =, \neq, \ge, >\}$ and the logical operators are given by $\{\neg,\wedge,\vee,\Rightarrow\}$, denoting negation, conjunction, disjunction, and implication, respectively. We use $\boolf$ to denote the obtained Boolean expressions, which are evaluated with respect to a given valuation $v \colon \boolf \to \{\vf, \vt\}$, where $\vf$ denotes the logical value false, and $\vt$ the logican value true. To this end, we update variables by a partial variable update function $f \colon \Svar \rightharpoonup \Sdat$. The updating of variables is coupled with the action transitions that are labeled by actions from the set~$\Sact$. The set $\Sact$ is formed by all possible communication actions over a set of channels $\Schan$, i.e, $\Sact \triangleq \{c!_m?_n \mid m, n \in \bbbn, c \in \Schan\}$. We write $c!_n$ for $c!_n?_0$ and $c?_n$ for $c!_0?_n$ for $n \in \bbbn$ and $c \in \Schan$, and we write $c!$ for $c!_1$ and $c?$ for $c?_1$. The set of process terms~$\Cterms$ is induced by $T$, given by:
\[
T :: = \dl \mid \st \mid a[f] . T \mid \gc{\phi}{T} \mid \encap{H}{T} \mid T + T \mid T \cdot T \mid T^* \mid \, T \pc T \
\]
where $a \in \Sact$, $f \colon \Svar \rightharpoonup \Sexp$, $\phi \in \boolf$, and $H \subseteq \Sact$. Each process $p \in \Cterms$ is coupled with a global variable assignment environment that is used to evaluate the guards and keeps track of updated variables, notation $\val{p}{(\alpha, \rho)} \in \Cterms \times \Sigma$ for $\Sigma = (\Svar \to \Sexp) \times \Svar$. By $\alpha\colon \Svar \to \Sexp$ we denote the assignment of the variables in order to consistently evaluate the guards, whereas the predicate $\rho \subseteq \Svar$ keeps track of the updated variables, which is needed for correct synchronization. We write $\sigma = (\alpha, \rho)$ for $\sigma \in \Sigma$, when the components of the environment are not explicitly required. The initial assignment $\sigma_0 = (\alpha_0, \dom(\alpha_0))$, where $\dom(f)$ denotes the domain of the function $f$, provides the initial values of all variables that the process comprises.

The theory has two constants: $\dl$ denotes deadlock that cannot execute any action, whereas $\st$ denotes the option to successfully terminate. The action-prefixed process with variable update, corresponding to $a[f] . p$, executes the action~$a$, while updating the data values according to $f$, and continues behaving as~$p$. The guarded command, notation $\gc{\phi}{p}$, specifies a guard $\phi \in \boolf$ that guards a process $p \in \Cterms$. If the guard is successfully evaluated, the process continues behaving as $p \in \Cterms$ or, else, it deadlocks. The encapsulation operator $\encap{H}{p}$ encapsulates the process $p$ in such a way that all communication actions in $H$ that are considered as incomplete are blocked, so that the desired type of communication is enforced. For example, if we were to enforce communication between $k$ processes over channel $c$, then $H = \{c!_m?_n \mid 0 < m + n, \ m + n \neq k\}$. The sequential composition $p \cdot q$ executes an action of the first process, or if the first process successfully terminates, it continues to behave as the second. The unary operator $p^{*}$ represents iteration, or the Kleene star, that unfolds with respect to the sequential composition. The alternative composition $p + q$ makes a nondeterministic choice by executing an action of~$p$ or~$q$, and continues to behave as the remainder of the chosen process. The binary operator $p \pc q$ denotes parallel composition with generic communication actions, where the actions of both arguments can always be interleaved or, alternatively, communication can take place over common channels, keeping track of the number of involved sender and receiver parties.
\begin{figure*}[t!]
\centering
\[\small
\begin{array}{c}

  \sos{}{\val{\emp}{\sigma} \downarrow} \quad

  \sos{}{\val{a[f].p}{(\alpha,\rho)} \tr{a} \val{p}{(\alpha\{\{X \mathord{\mapsto} e(f(X)) \mid X \mathord{\in} \dom(f)\}\}, \dom(f))}}

\quad

  \sossym{\val{p}{\sigma}\downarrow}{\val{p + q}{\sigma}\downarrow}
\smallskip \\
  \sossym{\val{p}{\sigma} \tr{a} \val{p'}{\sigma'}}
  {\val{p + q}{\sigma} \tr{a} \val{p'}{\sigma'}}
\quad
  \sos{\val{p}{\sigma}\downarrow, \ \val{q}{\sigma}\downarrow}
  {\val{p \cdot q}{\sigma}\downarrow}
\quad
  \sos{\val{p}{\sigma}\downarrow,\ \val{q}{\sigma} \tr{a} \val{q'}{\sigma'}}
  {\val{p \cdot q}{\sigma} \tr{a} \val{q'}{\sigma'}}
\quad
  \sos{\val{p}{\sigma} \tr{a} \val{p'}{\sigma'}}
  {\val{p \cdot q}{\sigma} \tr{a} \val{p' \cdot q}{\sigma'}}
\smallskip \\

  \sos{}{\val{p^*}{\sigma}\downarrow} \quad
  \sos{\val{p}{\sigma} \tr{a} \val{p'}{\sigma'}}{\val{p^*}{\sigma} \tr{a} \val{p'\cdot p^*}{\sigma'}}

\quad

  \sos{\val{p}{\sigma} \downarrow, \ \val{q}{\sigma}\downarrow}
  {\val{p \pc q}{\sigma}\downarrow}
\quad
  \sossym{\val{p}{\sigma} \tr{a} \val{p'}{\sigma'}}
  {\val{p \pc q}{\sigma} \tr{a} \val{p' \pc q}{\sigma'}}
  \smallskip \\
  \sos{
  \val{p}{\sigma} \tr{c!_k?_\ell} \val{p'}{(\alpha', \rho')}, \ \val{q}{\sigma} \tr{c!_m?_n} \val{q'}{(\alpha'', \rho'')}, \  \alpha'|_{\rho' \cap \rho''} = \alpha''|_{\rho' \cap \rho''}
}
  {\val{p \pc q}{\sigma} \tr{c!_{k+m}?_{\ell+n}} \val{p' \pc q'}{(\alpha'\{\alpha''|_{\rho'' \setminus \rho'}\}, \rho'\cup \rho'')}} \quad
\quad
  \sos{\val{p}{\sigma} \downarrow, \ v(\phi) = \vt}
  {\val{\gc{\phi}{p}}{\sigma}\downarrow}  \smallskip \\

  \sos{\val{p}{\sigma} \tr{a} \val{p'}{\sigma'}, \ v(\phi) = \vt}
  {\val{\gc{\phi}{p}}{\sigma} \tr{a} \val{p'}{\sigma'}}
\quad  \sos{\val{p}{\sigma} \downarrow}{\val{\encap{H}{p}}{\sigma} \downarrow}
   \qquad
      \sos{\val{p}{\sigma} \tr{a} \val{p'}{\sigma'}, \ 
              a \not \in H}%
           {\val{\encap{H}{p}}{\sigma} \tr{a}
             \val{\encap{H}{p'}}{\sigma'}}
\end{array}
\]
\caption{Operational rules}
\label{fig:operationalrules}
\end{figure*}

\subsection{Structural Operational Semantics}

We give semantics in terms of labeled transition systems coupled with a environment that keeps track of the valuation of the data variables and the updated variables. The states of the labeled transition systems are labeled by the process terms themselves, and the dynamics of the process is given by a successful termination option predicate $\tss{} \subseteq \Cterms \times \Sigma$, that plays the role of final or marked states for nonblocking supervision~\cite{rwsupervisor,Cassandras}, and an action transition relation $\tr{} \subseteq (\Cterms \times \Sigma) \times \Sact \times (\Cterms \times \Sigma)$. We write $\tss{\val{p}{\sigma}}$ for $\val{p}{\sigma} \in \tss{}$ and $\val{p}{\sigma} \tr{a} \val{p'}{\sigma'}$ for $(\val{p}{\sigma},a,\val{p'}{\sigma'}) \in \tr{}$.

To present concisely the update of the assignments, we introduce several auxiliary operations. We write $f|_C$ for the restriction of the function $f$ to the domain $C \subseteq \dom(f)$, i.e., $f|_C \triangleq \{x \mapsto f(x) \mid x \in C\}$. Also, we introduce the notation $f\{f_1\}\ldots\{f_n\}$, where $f \colon A \to B$ and $f_i\colon A \rightharpoonup B$ for $1 \le i \le n$ are partial functions with mutually disjoint domains, i.e., $\dom(f_i) \cap \dom(f_j) = \emptyset$ for $i \neq j$. For every $x \in A$, we have that $f\{f_1\}\ldots\{f_n\}(x) = f_j(x)$, if there exists some $j$ such that $1\le j \le n$ and $x \in \dom(f_j)$, or $f\{f_1\}\ldots\{f_n\}(x) = f(x)$ otherwise. We define $\tss{}$ and $\tr{}$ using structural operational semantics~\cite{pabook}, depicted by the operational rules in Figure~\ref{fig:operationalrules}. We note that symmetrical rules are not depicted, and their number is only denoted in brackets next to the number of the rule that is to be applied for the process on the left side of the operation.

Rule~1 states that the termination constant has the option to successfully terminate. Rule~2 states that the action prefix enables action transitions, whereas the target assignment must be updated accordingly. Namely, the variables in the domain of the partial variable assignment function are updated with the evaluated data expression. Rules 3 and 4 state that the alternative composition can successfully terminate if one of its summands has the option to successfully terminate. Similarly, action transitions are possible in the alternative composition if one of its summands can perform them, as given by rules 5 and 6. Rule 7 states that the sequential composition has a termination option if both of its components have a termination option. If the first component terminates, then the sequential composition continues behaving as the second component, as given by rule 8. If the first component can perform an action transition, then the target process sequentially composes the target of the action transition of the first component with the second component, as given by rule 9. Iteration always has a termination option as given by rule 10, because of properties of composition of recursive processes~\cite{pabook}. Rule 11 shows that iteration unfolds with respect to sequential composition. Rule 12 states that the parallel composition can successfully terminate only if both of its components have successful termination options. Rules 13 and 14 enable interleaving in the parallel composition, even interleaving of transitions that stem from the same channel. Synchronizing of action transitions is possible for action that stem from the same channel as depicted by rule 15. The resulting communication action must account for the accumulative number of sender and receiver communication parties. The sets $\rho'$ and $\rho''$ identify the updated variables, so the common updated variables are given by $\rho' \cap \rho''$. The target environment updates the target environment of the first component with the remaining target environment of the second component, which can also be done symmetrically with respect to the second component. Rules 16 and 17 state that if the propositional guard is successfully evaluated, then the guarded command can successfully terminate or perform an action transition, respectively, provided that the guarded process can do so. Rule 18 states that successful termination is not affected by the encapsulation operator, whereas rule 19 states that all actions in the parameter set $H \subseteq \Sact$ are blocked.

\subsection{Partial Bisimulation}

The behavioral relation that we employ is an extension of partial bisimulation~\cite{acc2011}, which is able to handle data and variable assignments. Here, we directly employ the approach of~\cite{signalsBaeten,pabook}, where this extension is shown for bisimulation.

A relation~$R \subseteq \Cterms \times \Cterms$ is said to be a partial bisimulation with respect to a bisimulation action set~$B \subseteq \Sact$, if for all $(p,q) \in R$ and $\sigma \in \Sigma$, it holds that:
\begin{enumerate}
\item  $\val{p}{\sigma}\downarrow$ 
if and only if $\val{q}{\sigma}\downarrow$;

\item 
if $\val{p}{\sigma} \tr{a} \val{p'}{\sigma'}$ for $a \in \Sact$, then there exist $q' \in \Cterms$ and $\sigma' \in \Sigma$ such that $\val{q}{\sigma} \tr{a} \val{q'}{\sigma'}$ and $(p',q') \in R$;

\item 
if $\val{q}{\sigma} \tr{b} \val{q'}{\sigma'}$ for $b \in B$, then there exist $p' \in \Cterms$ and $\sigma' \in \Sigma$ such that $\val{p}{\sigma} \tr{b} \val{p'}{\sigma'}$ and $(p',q') \in R$.
\end{enumerate}
If $R$ is a partial bisimulation relation such that $(p, q) \in R$, then $p$ is partially bisimilar to $q$ with respect to $B$ and we write $p \pbis{B} q$. If $q \pbis{B} p$ holds as well, we write $p \bpbis{B} q$.

It is not difficult to show that partial bisimilarity is a preorder for the process terms in $\Cterms$~\cite{reportPACO} following the guidelines of~\cite{signalsBaeten}. In addition, following the guidelines of~\cite{coalgebra}, it can be shown that $\pbis{B}$ is a partial bisimulation relation with respect to $B \subseteq \Sact$. Thus, we obtain the partial bisimulation preorder and equivalence, similarly as for simulation preorder and equivalence~\cite{glabbeek}. Moreover, the partial bisimulation preorder can be shown a precongruence for the considered processes operations following the guidelines of~\cite{reportPACO,acc2011}, where a suitable extension to the tyft format for structural operational semantics with data of~\cite{tyftdata} is proposed. Consequently, the partial bisimulation equivalence is a congruence, which enables us to build a standard term model using the quotient algebra modulo $\bpbis{B}$ in the vein of~\cite{pabook}. Finally, we note that $p \bpbis{\Sact} q$ amounts to bisimulation~\cite{acc2011}, whereas $p \pbis{\emptyset} q$ reduces to simulation preorder~\cite{acc2011} and $p \bpbis{\emptyset} q$ reduces to simulation equivalence~\cite{acc2011} for processes with data~\cite{pabook,glabbeek}.

\section{A Process-Theoretic Approach to Supervisory Coordination}

First, we characterize the process terms that can be used to specify the plant and the supervisor. Thus, we distinguish between the two different flows of information on syntactic level. We employ the notion of partial bisimulation to define the relationship between the plant and the supervisor in order to ensure that the supervisor does not disable any uncontrollable events. Thereafter, we identify a set of data-based control requirements that are typically employed in specification documents. Finally, we describe the model-based system engineering framework and we discuss its implementation.

\subsection{Plant and Supervisor Syntax}

We distinguish between controllable and uncontrollable actions transitions that stem from the sets of controllable $\Schan_\Cact$ and uncontrollable $\Schan_\Uact$ channels, where $\Schan_\Cact \cap \Schan_\Uact = \emptyset$ and $\Schan_\Cact \cup \Schan_\Uact = \Schan$. We put $\Cact \triangleq \{c!_m?_n \mid m, n \in \bbbn, \ c \in \Schan_\Cact\}$ and $\Uact \triangleq \{u!_m?_n \mid m, n \in \bbbn, \ u \in \Schan_\Uact\}$. We model marked states by adding a successfully termination option to the corresponding state. We note that in the process theoretic setting, successful termination plays an additional role of enabling the sequential composition of processes~\cite{pabook,PACO2011}, which is not present in the automata theory of~\cite{rwsupervisor,Cassandras}. We restrict the syntax of the plant and the supervisor, given by $P$ and $S$, respectively, as follows:
\begin{equation}\label{eq:plantsyntax}
\begin{array}{l}
P :: = \dl \mid \st \mid c?_n[f] . P \mid u!_m?_n[f] . P \mid \gc{\phi}{P} \mid \encap{H}{P} \mid P + P \mid P \cdot P \mid P^* \mid \, P \pc P \ \smallskip \\
S :: = \st \mid c![\emptyset] . S \mid S + S \mid \gc{\phi}{S} \mid S^*,
\end{array}
\end{equation}
where $c \in \Schan_\Cact$, $u \in \Schan_\Uact$, $f \colon \Svar \rightharpoonup \Sexp$, $m, n \in \bbbn$, $\phi \in \boolf$, and $H \subseteq \Sact$. We note that we specify a monolithic supervisor, i.e., the supervision is executed by a single process. For modular or distributed supervision~\cite{Cassandras}, the syntactic form of the supervisor from~(\ref{eq:plantsyntax}) should be adjusted appropriately, so that it can admit several concurrent communicating processes.

We require the supervisor to be a deterministic process~\cite{acc2011}, which sends feedback to the plant in terms of synchronizing controllable events, and it does not alter the state of the plant in any other way, i.e., it comprises no variable assignments. The supervisor relies on data observation from the plant to make supervision decisions in the vein of~\cite{extendedguards}. Thus, the supervisor observes the state of the plant, identified by the values of the (shared) variables, and enables controllable events by synchronizing with a corresponding sender event. It does not influence uncontrollable events, so they are safely interleaved in the communication with the plant. Consequently, the supervisor does not have to keep a history of events, so it can be also be defined as an iterative process, which observes assigned data by employing guarded commands. This alternative definition is given as:
\begin{equation} \label{eqn:statebasedsupervisor}
s = \big(\textstyle \sum_{c \in \Schan_\Cact} \gc{\phi_c}{c![\emptyset].1} + \gc{\psi}{1}\big)^*,
\end{equation}
where $\phi_c, \psi \in \boolf$ for $c \in \Schan_\Cact$. A supervisor of form (\ref{eqn:statebasedsupervisor}) employs data value observation to identify the state of the plant and send back feedback regarding controllable events by synchronizing on self loops, as specified by $\sum_{c \in \Schan_\Cact} \gc{\phi_c}{c![\emptyset].1}$. It can potentially disable undesired termination options in states identified by $\psi \in \boolf$. The guards $\phi_c$ for $c \in \Schan_\Cact$ and $\psi$ depict the supervision actions~\cite{extendedguards}.

\subsection{Supervised Plants and Controllability}

If we suppose that the plant is given by $p \in P$ and the supervisor is given by $s \in S$, then the supervised plant can be specified as $\encap{H}{p \pc s}$ in general, where the encapsulation enforces desired communication and the set $H \subset \Sact$ comprises unfinished communication events, which differ per case. To ensure that no uncontrollable events are disabled by the supervisor, we employ partial bisimilarity to provide a relation between the supervised and the original plant. We note, however, that most of the other approaches, like~\cite{rwsupervisor,Cassandras,heymann-nondeterministic,kumarnondeterministic} to name a few, employ synchronizing actions, where two transitions with the same label synchronize in a resulting transitions, which is again labeled by that same label. In that case, the relation can be provided directly as in~\cite{acc2011}, because the labels of the transitions in the supervised and the original plant coincide.

\begin{figure}
\[
\begin{array}{c}
\sos{\tss{\val{p}{\sigma}}}{\tss{\val{\xi(p)}{\sigma}}} \qquad \sos{\val{p}{\sigma} \tr{c?_n[f]} \val{p'}{\sigma'}, \ c \in \Schan_\Cact}{\val{\xi(p)}{\sigma} \tr{c!?_n[f]} \val{\xi(p')}{\sigma'}} \qquad \sos{\val{p}{\sigma} \tr{u!_m?_n[f]} \val{p'}{\sigma'}, \ u \in \Schan_\Uact}{\val{\xi(p)}{\sigma} \tr{u!_m?_n[f]} \val{\xi(p')}{\sigma'}}
\end{array}
\]
\caption{Renaming operation that renders the controllable communication events of the plant completed}\label{fig:renaming}
\end{figure}

In the setting of this paper, however, we have to rename certain actions in the original plant so that we can mimic the presence of a supervisor, which is necessitated in order to make the plant operational. To this end, we employ a specific partial renaming operation $\xi \colon \Cterms \mapsto \Cterms$ that renders the controllable communication actions of the plant as completed. This is in accordance with the syntax of the plant and the supervisor specified in~(\ref{eq:plantsyntax}), since the plant must wait for an enabling control signal for every controllable event. The operational rules that define the renaming operation $\xi$ are given in Figure~\ref{fig:renaming}.

Now, we can specify the relation between the supervised and the original plant as:
\begin{equation}\label{eqn:controllability}
\encap{H}{p \pc s} \pbis{\Uact} \xi(p).
\end{equation}
It states that the supervised plant has controllable events enabled by the supervisor that can be simulated by the original plant in which all controllable events have been enabled, whereas no uncontrollable events can be disabled. We note that, in the setting of this paper, one can observe that the syntactical restrictions imposed on the supervisor actually imply this relation.

It is not difficult to show, again in the vein of~\cite{coalgebra,acc2011,reportPACO}, that the traditional notions of language-based controllability of~\cite{rwsupervisor,Cassandras} for deterministic system and state controllability~\cite{extendedguards,fabiannondeterministic,kumarnondeterministic} for nondeterministic systems are implied by~(\ref{eqn:controllability}).

\subsection{Data-Based Control and Coordination Requirements}

In the setting of this paper, we consider data-based control and coordination requirements, which are stated in terms of boolean expressions ranging over the data variables, and may additionally specify which events are allowed with respect to the observed data values. For a setting with event-based control requirements, we refer the interested reader to~\cite{acc2011}, whereas for state-based control requirements, a preliminary investigation is given in~\cite{PACO2011}. The data-based control requirements, denoted by the set $\Sreq$, have the following syntax induced by $R$:
\[
R ::=\ \tr{a} \Rightarrow \phi \ \mid\  \phi \Rightarrow \ntr{a} {} \  \mid \  \phi,
\]
for $a \in \Sact$ and $\phi \in \boolf$. A given control requirement $r \in \Sreq$ is satisfied with respect to the root of the process term $p \in \Cterms$ in the assignment environment $\sigma \in \Sigma$, notation $\val{p}{\sigma} \models r$, according to the operational rules depicted in Figure~\ref{fig:statebasedrequirements}. By $\val{p}{\sigma} \ntr{a} {}$ we denote that $\{\val{p'}{\sigma'} \mid \val{p}{\sigma} \tr{a} \val{p'}{\sigma'}\}= \emptyset$.

\begin{figure}[!tb]
\[
\begin{array}{c}
\sos{\val{p}{\sigma} \models \neg \phi \Rightarrow \ntr{a}}{\val{p}{\sigma} \models \tr{a} \Rightarrow \phi} \quad
\sos{v(\phi) = \vf}{\val{p}{\sigma} \models \phi \Rightarrow \ntr{a}}
\quad
\sos{\val{p}{\sigma}\ntr{a} {}}{\val{p}{\sigma} \models \phi \Rightarrow \ntr{a}} \quad \sos{v(\phi) = \vt}{\val{p}{\sigma} \models \phi}
\end{array}
\]
\caption{Satisfiability of data-based control requirements}\label{fig:statebasedrequirements}
\end{figure}

The first form of control requirements is introduced for modeling convenience as a frequently occurring case~\cite{CDC2010} and it is equivalent to the second form, as given by rule~23. Rule~24 states that if the state does satisfy the data assignment, then the requirement is trivially satisfied. Rule~25 states a so-called state-transition exclusion requirement~\cite{CDC2010}, which is satisfied if no transition with the excluded label is possible. Rule~26 states that a state-exclusion requirement restricts the states with the given data assignments, thus disabling unsafe or forbidden states, and must be upheld in every state. To ensure that the control requirements are globally satisfied, we extend $\models$ to $\models^*$, which requires that the control requirements are satisfied for every reachable state. To this end, we first define a trace transition relation $\val{p}{\sigma} \trs{t} \val{p'}{\sigma'}$ for some $t = a_1 \ldots a_n \in \Sact^*$. If $n = 0$ we have the empty trace $t = \epsilon$ with $\val{p}{\sigma} \trs{\epsilon} \val{p}{\sigma}$, whereas if $n > 0$, then we have $\val{p}{\sigma} \tr{a_1} \val{p_1}{\sigma_1} \tr{a_2} \val{p_2}{\sigma_2} \tr{a_3}\ldots \tr{a_n} \val{p'}{\sigma'}$ for some $p_1, \ldots, p_{n-1} \in \Cterms$, $\sigma_1, \ldots, \sigma_{n-1} \in \Sigma$, and $a_1, \ldots, a_{n-1} \in \Sact$.  Now, we define that $p \models^* r$ if $q \models r$ for every $p' \in \Cterms$ such that $\val{p}{\sigma} \trs{t} \val{p'}{\sigma'}$ for $\sigma, \sigma' \in \Sigma$ and $t \in \Sact^*$.

To ensure that the supervised plant respects the data-based control requirements, given by $R \subset \Sreq$, we require that for the initial variable assignment $\sigma_0 \in \Sigma$ it holds that
\begin{equation}\label{eqn:controlrequirements}
\textstyle \val{p \pc s}{\sigma_0} \models^* \bigwedge_{r \in C} r.
\end{equation}
In addition, a nonblocking supervisor must ensure that every state in the supervised plant can reach a state that has a successful termination option, i.e., for every $\val{p'}{\sigma'} \in \Cterms \times \Sigma$ and $t \in \Sact^*$ such that $\val{p \pc s}{\sigma_0} \trs{t} \val{p'}{\sigma'}$, there exists $\val{p''}{\sigma''} \in \Cterms \times \Sigma$ and $t' \in \Sact^*$ such that $\val{p'}{\sigma'} \trs{t'} \val{p''}{\sigma''}$ and $\tss{\val{p''}{\sigma''}}$ holds.


\begin{figure}[!t]
  \centering
  \includegraphics[width=\textwidth]{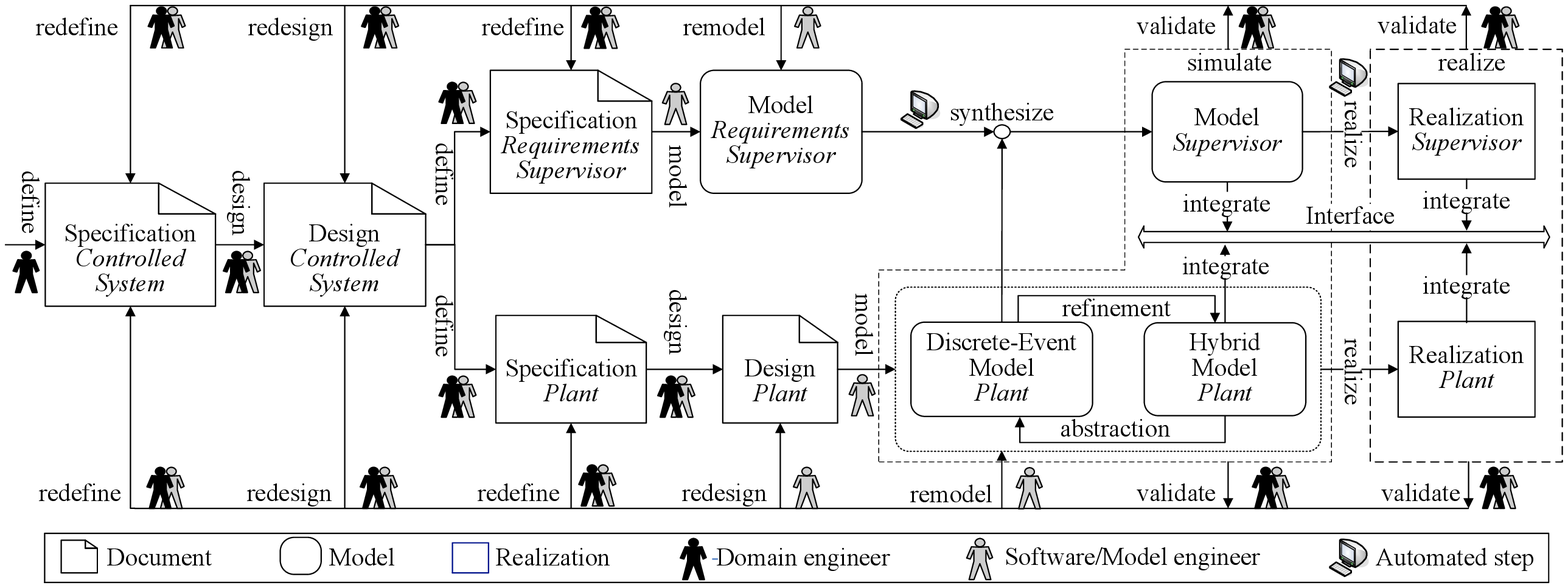}\\
  \caption{Model-based systems engineering framework for supervisory controller synthesis}\label{fig:framework}
\end{figure}

\subsection{Model-Based Systems Engineering Framework}

To structure the process of supervisory control synthesis we employ the framework depicted in Figure~\ref{fig:framework}~\cite{mbescs,CDC2010,cosy2011}. The modeling process begins with an informal specification of the controlled system, i.e., the desired product, written by domain engineers. A design of the controlled system follows, contrived by domain and software engineers together. The design most importantly defines the modeling level of abstraction and the control architecture. Subsequently, it is used to separate the plant and the control requirements, a joint task of domain and software engineers. Here, a decision is made to which extent the control is managed by the software, and which part is implemented in hardware. The resulting informal documents specify the plant and control requirements, respectively. In the following, we omit the roles of the engineers as they are clear from the context.

Most plants typically contain (continuous) hybrid behavior, whereas supervisor synthesis requires a discrete-event abstraction. The hybrid model is suitable for simulation purposes, and it can be abstracted to a discrete-event model for synthesis purposes~\cite{rwsupervisor,Cassandras}. Alternatively, a discrete-event model can be made, and subsequently refined~\cite{tabuadalinear}. In the design of the plant, decisions are made on the level of abstraction that is used, and what is significant discrete-event and hybrid behavior. In parallel, a model of the control requirements is made following the specification documents. The discrete-event model of the plant, together with the model of the control requirements, are input to the synthesis tool, which automatically synthesizes a supervisor.

Software-in-the-loop simulation is used to validate the supervisor coupled with a hybrid model of the plant, and hardware-in-the-loop simulation can be used to validate the supervisor against a prototype of the plant. If the validation is not satisfactory, the control requirements and/or the plant model need to be remodeled or redefined. In certain cases, a complete revision proves to be necessary, which might even require redefining the specification of the whole controlled system. Finally, the control software is generated automatically, based on the validated models. Note that software engineers in the framework act more as `model' engineers, shifting their focus from writing code to modeling.

We opt for Supremica~\cite{supremica} as a synthesis tool because it provides the greatest modeling convenience and range of options with respect to specifying plants with data and optimized synthesis procedures~\cite{extendedguards}. We remark that the state-of-the-art synthesis tools support the prevailing automata-style specifications and composition~\cite{supremica,Cassandras}. To be able to execute industrial case studies, we are impelled to translate the original process-algebraic specification to an input accepted by the tool.

\newcommand{\m}[1]{\mathit{#1}}
\newcommand{\mr}[1]{\mathrm{#1}}
\newcommand{\mto}{\mapsto}

\section{Coordinating Maintenance Procedures of a Printing Process Function}

\begin{figure}[!t]
  \centering
  \includegraphics[width=0.6\textwidth]{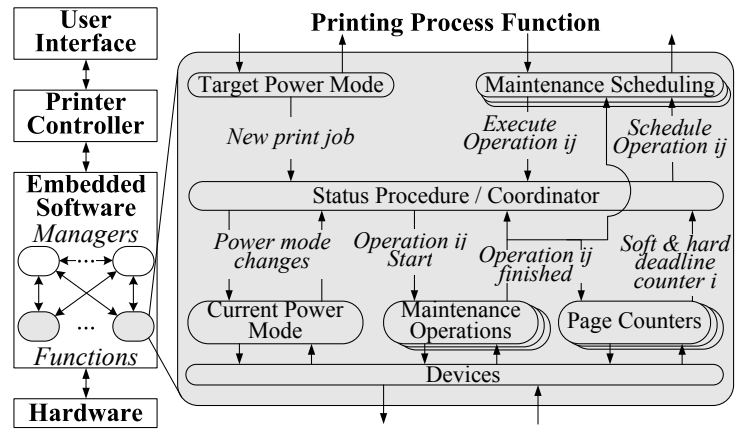} 
\caption{Modeling of the printing process function}\label{fig:detailedfunction}
\end{figure}

An abstract view of the control architecture of a high-tech printer is depicted in Figure~\ref{fig:detailedfunction}. Print jobs are sent to the printer by means of the user interface. The printer controller communicates with the user and assigns print jobs to the embedded software, which actuates the hardware to realize print jobs. The embedded software is organized in a distributed way, per functional aspect, such as, paper path, printing process, etc. Several managers communicate with the printer controller and each other to assign tasks to functions, which take care of the functional aspects.

We depict a printing process function comprising several maintenance operations in Figure~\ref{fig:detailedfunction}. Each function is hierarchically organized to: (1) controllers: Target Power Mode and Maintenance Scheduling, which receive control and scheduling tasks from the managers; (2) procedures: Status Procedure, Current Power Mode, Maintenance Operation, and Page Counter, which handle specific tasks and actuate devices, and (3) devices as hardware interface. Status Procedure is responsible for coordinating the other procedures given the input form the controllers. The control problem is to synthesize a supervisory coordinator that ensures that quality of printing is not compromised by timely performing maintenance procedures, while interrupting ongoing print jobs as little as possible. We specify the coordination rules that ensure safe behavior of the system below.

\begin{figure}[!t]
\[
\begin{array}{@{}l}

{CPM} \triangleq \big(\vev{Stb2Run?}\{\m{CPM \mto 2}\}. \vev{\_InRun}\{\m{CPM \mto 3}\}.
\vev{Run2Stb?}\{\m{CPM \mto 4}\}. \vev{\_InStb}\{\m{CPM \mto 1}\}.1 + 1\big)^*

\smallskip \\

\textstyle {MO_{ij}} \triangleq \big(\vev{OpStart_{ij}}\{\m{MO_{ij} \mto 2}\}. \vev{\_OpFin_{i}!}\{\m{MO_{ij} \mto 1}\}.1 + 1\big)^* \smallskip \\

{PC_i} \triangleq \big(\vev{\_SoftDln_i}\{\m{PC_i \mto 2}\}. \big( \vev{\_HardDln_i}\{\m{PC_i \mto 3}\}. \vev{\_OpFin_{i}?}\{\m{PC_i \mto 1}\}.1 +
\vev{\_OpFin_{i}?}\{\m{PC_i \mto 1}\}.1\big) + {} \smallskip \\
\phantom{{PC_i} \triangleq \big(}
\vev{\_OpFin_{i}?}.1 + 1\big)^* 
\smallskip \\

\textstyle {MS_i} \triangleq \big(\vev{SchOper_i?}\{\m{MS_i \mto 2}\}. \vev{\_ExOper_i}\{\m{MS_i \mto 3}\}. 
\vev{\_OpFin_{i}?}\{\m{MS_i \mto 1}\}.1 + 1\big)^*
\smallskip \\

\textstyle {TPM} \triangleq \big(\vev{\_NewJob}\{\m{TPM \mto 2}\}. 
\vev{\_JobFin}\{\m{TPM \mto 1}\}.1 + 1\big)^* \smallskip \\

{PPF} \triangleq \encap{\{\_OpFin_{i}?, \_OpFin_{i}?_2, \_OpFin_{i}!?\}}{{CPM} \pc \big( \mbox{\large $\pc$} _{i \in I} {MS_i} \big) \pc \big( \mbox{\large $\pc$} _{j \in J_i,\, i \in I} {MO_{ij}} \big) \pc \big( \mbox{\large $\pc$} _{i \in I} {PC_i} \big) \pc {TPM}}

\end{array}
\]
\caption{Process-algebraic specification of the plant}\label{fig:plant}
\end{figure}

\subsection{Process-Algebraic Specification}

We briefly describe the procedures that comprise the plant, whose process-algebraic specification is given in Figure~\ref{fig:plant}. We assume that the page counters are indexed by the set $I$, whereas for each counter $i$ there are $J_i$ maintenance procedures to be triggered. Also, labels of uncontrollable events begin with an underscore. Furthermore, we identify states by means of variable observation, i.e., we incorporate the observer inside the plant specification, so we assign the variables $\mr{MO}_{ij}$ to Maintenance Operation $ij$, $\mr{PO}_i$ to Page Counter $i$, $\mr{MS}_i$ to Maintenance Scheduling $i$, $\mr{TPM}$ to Target Power Mode, and $\mr{CPM}$ to Current Power Mode for $j \in J_i$ and $i \in I$. Initially, the variables are set to $1$, which identifies the first state. The plant model is depicted in Figure~\ref{fig:plant}, where Printing Process Function is defined by $\mathrm{PPF}$ and $\mbox{\large $\pc$}_{p \in P} p$ denotes the parallel composition of the processes in $P$.

Current Power Mode sets the power mode to run or standby depending on the enabling signals (\emph{Stb2Run} and \emph{Run2Stb}) from Status Procedure, and sends back feedback by employing \emph{\_InRun} and \emph{\_InStb}, respectively. Maintenance Operation $ij$ for $j \in J_i$ and $i \in I$ either carries out a maintenance operation, started by $\_\m{OpStart}_{ij}$ or it is idle. The confirmation is sent back by the events $\vev{\_OpFin}_{ij}!$ for $j \in J_i$ and $i \in I$, which synchronize with Maintenance Scheduling and Page Counter. Page Counter $i$ for $i \in I$ counts the printed pages since the last maintenance and sends signals $\m{\_SoftDln}_{i}$ and $\m{\_HardDln}_{i}$, when soft or hard deadlines are reached, respectively. It is responsible for the set of maintenance procedures in $J_i$. A soft deadline signals that maintenance should be performed, but it is not yet compulsory if there are pending print jobs. A hard deadline is reached when maintenance of the printing process must be performed to ensure quality of the print. The page counter is reset, triggered by the synchronization on $\vev{\_OpFin}_{ij}?$, each time that maintenance is finished. The controller Target Power Mode sends signals regarding incoming print jobs to Status Procedure by \emph{\_NewJob}, which should set the printing process to run mode for printing and standby mode for maintenance and power saving. When the print job is finished, the signal \emph{\_NoJob} is sent. Maintenance Scheduling $i$ for $i \in I$ receives a request for maintenance with respect to expiration of Page Counter $i$ from Status Procedure, by the signal $\m{SchOper_i}$ and forwards it to the manager. The manager confirms the scheduling with the other functions and sends a response back to the Status Procedure, using $\m{\_ExOper_i}$. It also receives feedback from Maintenance Operation that the maintenance is finished in order to reset the scheduling, again triggered by $\vev{\_OpFin}_{ij}?$.

\subsection{Coordination Requirements}

Status Procedure adheres to several coordination rules:

1) \emph{Maintenance operations can be performed only when Printing Process Function is in standby.} This state exclusion property requires a maintenance operation $ij$ to be in progress, identified by $\m{MO}_{ij} = 2$, only if the printer is in standby, i.e., $\m{CPM} = 1$. Thus, we specify that the following must always hold:
\begin{equation}
\textstyle \neg (\m{CPM} \neq 1 \wedge \bigvee_{i \in I, j \in J_i} \m{MO}_{ij} = 2)
\end{equation}

2) \emph{Maintenance operations can be scheduled only if soft deadline has been reached and there are no print jobs in progress, or a hard deadline is passed.} We schedule a maintenance operation $ij$ for $j \in J_i$ using the signal $\mathit{SchOper_i}$ for $i \in I$. Soft and a hard deadline for Page Counter $i$ is identified by $\m{PC_i} = 2$ and $\m{PC_i} = 3$, respectively, leading to
\begin{equation}
\tr{\vev{SchOper_i!?}} \Rightarrow ( \m{PC_i} = 2 \wedge \m{TPM} = 1) \vee \m{PC_i} = 3
\end{equation}
for every $i \in I$.

3) \emph{Maintenance operations can be started only after being scheduled.} For every $j \in J_i$ and $i \in I$. Thus, we relate $\vev{OpStart_{ij}}$ with the corresponding maintenance scheduler:
\begin{equation}
\tr{\vev{OpStart_{ij}!?}} \Rightarrow \m{MS}_i = 3.
\end{equation}

4) \emph{The power mode of the printing process function must follow the power mode dictated by the managers, unless overridden by a pending maintenance operation.} We model this requirement separately for switching from run to standby power mode and vice versa. We can switch from run to standby if this is required by the manager, i.e., there is a new print job, and there is no need to start a maintenance operation. This is modeled as
\begin{equation}
\textstyle \tr{\mathit{Stb2Run}!?} \Rightarrow \m{TPM} = 2 \wedge \bigwedge_{i \in I} \m{MS}_i \neq 3.
\end{equation}
Contrariwise, we switch to Standby if there is no pending job or maintenance operation:
\begin{equation}
\textstyle \tr{\mathit{Run2Stb}!?} \Rightarrow \m{TPM} = 1 \vee \bigvee_{i \in I} \m{MS}_i = 3.
\end{equation}
The set of parameterized data-based coordination requirements is given by the expressions (5) -- (9).

\subsection{Supervisor Synthesis}

For any value of the parameters for the index sets $I$ and $J_i$ for $i \in I$, we can instantiate a plant and synthesize a supervisor. For the sake of clarity, we illustrate the situation when there is only one maintenance procedure. We omit the unnecessary indices of the data variables. The supervisor sends the control signals upon observation of certain data assignments, which are given in the form of guards. The indices of the guards correspond to the indices of the control requirements that concern the control signal. Note that the state-exclusion requirement is treated as a global invariant, whereas no termination option of the plant is disabled. The guards have been synthesized as follows~\cite{extendedguards}:
\[
\begin{array}{ll}

g_6 \triangleq (\m{PC = 2}\wedge \m{TPM = 1})\vee \m{PC = 3} \qquad & \qquad

g_7 \triangleq \m{CPM = 1} \wedge \m{MS = 3} \smallskip \\

g_8 \triangleq \m{MS \neq 3} \wedge \m{TPM = 2} \wedge \m{MO \neq 2} \qquad&\qquad

g_9 \triangleq (\m{MS \neq 3} \wedge \m{TPM = 1} )\vee \m{MS = 3}.
\end{array}
\]
The supervisor has the syntax form restricted by (1) and it is given by:
\[
\begin{array}{l}
S \triangleq \Big( \gc{g_6}{\vev{SchOper!}}.\emp + \gc{g_7}{\vev{OpStart!}}.\emp + \gc{g_8}{\vev{Stb2Run!}}. \emp + \gc{g_9}{\vev{Run2Stb!}}.\emp + 1\Big)^*.
\end{array}
\]

To illustrate the process of supervision, we consider the event $\vev{Stb2Run}$. It is not difficult to deduce, e.g., that initially the event \emph{Stb2Run} is not enabled since then all variables are assigned the value of $1$. This corresponds to the situation where there are not print jobs waiting to be executed, so there is no reason to turn the power of the printer on. Similarly, a maintenance operation can be started only if the printer is in standby mode, identified by $\m{CPM} = 1$, and the operation has been successfully scheduled, identified by $\m{MS} = 3$.

\section{Concluding Remarks}\label{sec:conclusion}

We developed a process theory encompassing communicating processes with data and generic communication actions. We applied the developed theory to model supervisory control feedback loops with data observations, where we distinguish between the observation and control flow of information. We classified the processes modeling the unsupervised system and the supervisory controller to capture their specific roles. To capture the notion of controllability, which identifies the set of feasible supervisory controllers, we employed the behavioral relation partial bisimulation and we extended the notion for the new setting. We casted the process of supervisory controller synthesis in a model-based systems engineering framework, for which implementation we employ state-of-the-art tools. To illustrate our approach, we reiterated on an industrial study dealing with coordination of maintenance procedures in a printing process of a high-tech printer. We demonstrated that our approach is capable of successfully modeling the interaction in the supervisory control loop and offers a compact representation of the model of the supervisory controller.


\bibliographystyle{eptcs}
\bibliography{bibliography}


\end{document}